\documentclass[
aip,
rsi,
amsmath,
amssymb,
reprint
]{revtex4-2}

\usepackage{graphicx}
\usepackage{bm}
\usepackage{booktabs}
\usepackage{hyperref}
\usepackage{booktabs}
\usepackage{multirow}
\usepackage{subcaption}
\usepackage{float}
\usepackage{xcolor}

\hypersetup{
	colorlinks = true,
	linkcolor = blue,
	citecolor = blue,
	urlcolor = blue
}

\begin{document}
	
	\title{A Signal Analysis Framework for Unshielded Room-Temperature Magnetocardiography}
	
	\author{Kushal Patel$^{\dagger}$}
	\author{Kesavaraja C$^{\dagger}$}
	\author{Pranab Dutta$^{\dagger}$}
	\email{pranab@gdqlabs.com}
	\author{Korak Biswas$^{\dagger}$}
	\affiliation{$^{\dagger}$GDQLABS Private Limited, Bengaluru, India}
	
	\date{\today} 
	
	\begin{abstract}
		Room-temperature, unshielded recording of cardiac magnetic signals has remained a significant challenge since the inception of magnetocardiography (MCG). In this work, we present an MCG system based on optically pumped magnetometers (OPMs) designed to operate in ambient magnetic environments and acquire adult human cardiac magnetic fields, without the need for active or passive shielding. The system operates in a gradiometer configuration, achieving background-noise cancellation with a common-mode rejection ratio (CMRR) of 31 dB and a gradient sensitivity of 314 $\mathrm{fT/cm/\sqrt{Hz}}$. MCG signals were acquired sequentially at 16 locations across the anterior thorax, and a comprehensive signal-analysis framework incorporating wavelet multiscale principal component analysis (WMSPCA) filtering and signal quality estimation (SQE) scoring was developed to enhance signal quality. This framework yielded a QRS complex signal-to-noise ratio (SNR) of 28.56 ± 5.61 dB across all measurement locations. These results demonstrate the feasibility of performing clinical-grade MCG in unshielded, real-world magnetic environments, with consistent morphological fidelity across the QRS complex and T-wave segments. This work represents a meaningful step toward the practical deployment of OPM-based MCG systems in hospital and point-of-care settings.
	\end{abstract}
	
	\maketitle
	
	\section{Introduction}
	
Magnetocardiography (MCG) is a non-invasive and non-contact technique used to measure the magnetic fields generated by cardiac electrical activity. Compared with electrocardiography (ECG), MCG provides improved spatial characterization of cardiac electrophysiological activity because it is less affected by distortions arising from tissue conductivity \cite{malmivuo1995bioelectromagnetism, roth2023biomagnetism}. These advantages make MCG a promising tool for diagnosing cardiac disorders such as coronary artery disease (CAD), left ventricular hypertrophy, cardiac arrhythmias, and myocarditis \cite{fenici2005clinical, chen2025new, yamada2005magnetocardiograms}.

Among the clinical applications of MCG, early-stage CAD detection has attracted considerable investigative attention. CAD represents the leading cause of cardiovascular mortality globally, and the identification of ischemia-induced electrophysiological abnormalities prior to the emergence of structural myocardial damage remains a critical diagnostic objective \cite{stark2024global}. MCG is particularly well-suited to this challenge, as it can resolve spatially heterogeneous repolarization disturbances and localized conduction delays electrical signatures of subclinical ischemia at stages when standard ECG indices remain within normal limits \cite{her2023magnetocardiography, li2026meta}. Prospective clinical evaluations have further demonstrated the translational potential of this capability, with MCG showing promise as a rapid, non-contact triage device in emergency department settings for the assessment of suspected acute coronary syndromes \cite{MACE2024100441}.

Recent advances in room-temperature magnetic sensing technologies have opened a new frontier in cardiac diagnostics, enabling MCG systems that operate without the cryogenic infrastructure and magnetically shielded rooms historically required by SQUID (superconducting quantum interference device)-based platforms \cite{clarke2006squid, sternickel2006biomagnetism}. Among the emerging sensor modalities including nitrogen-vacancy-center diamond sensors, magnetoresistive sensors, fluxgate sensors, induction-coil sensors, and tunnel magnetoresistance sensors, optically pumped magnetometers (OPMs) have demonstrated a particularly compelling combination of sensitivity and operational flexibility \cite{omar2026human, yaga2025recording, janosek2020fluxgate, mooney2017portable, xing2025multilevel, yang2021wearable, pena2020ninety}. When configured in a gradiometer arrangement, OPMs provide effective suppression of spatially correlated ambient magnetic noise, establishing a viable pathway toward reliable cardiac magnetic field measurement in unshielded clinical environments. Building on this foundation, the present work describes the development of an unshielded MCG system that directly addresses the challenge of ambient noise through a combination of hardware-level gradiometry and a robust signal processing framework designed to maintain signal quality and spatial consistency across measurement conditions \cite{iwata2024bedside, xiao2023movable, limes2020portable, patel2026magnetocardiography}.

In unshielded measurements, reliable reconstruction of cardiac magnetic field maps requires averaging a large number of cardiac cycles, making the recordings highly susceptible to motion artifacts, environmental interference, sensor drift, and polarity inconsistencies introduced during the process of denoising. Existing studies, including studies on shielded MCG, have largely focused on signal acquisition and noise suppression, while systematic approaches for beat-quality assessment remain limited \cite{lu2018realtime, jia2025novel, zhang2009quantitative, zhu2025multichannel, mariyappa2015denoising}.
	
\begin{figure*}[t]
	\centering
	
	\begin{subfigure}[t]{0.43\textwidth}
		\centering
		\includegraphics[width=\linewidth,width=\linewidth,trim=0cm 3.5cm 0cm 3.5cm,
		clip]{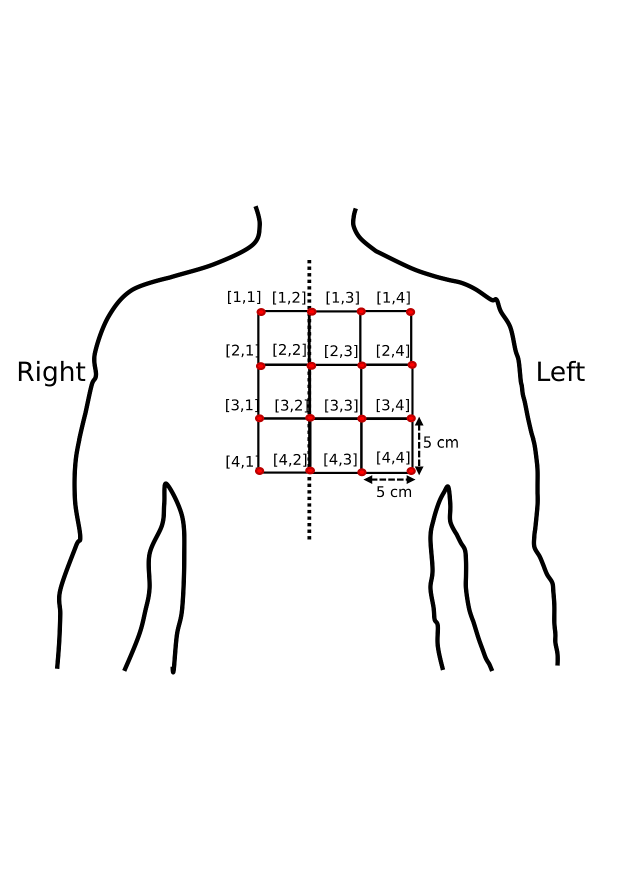}
		\caption{}
		\label{fig:grid}
	\end{subfigure}
	\hfill
	\begin{subfigure}[t]{0.48\textwidth}
		\centering
		\includegraphics[width=\linewidth,trim=1cm 4cm 0.1cm 4cm,
		clip]{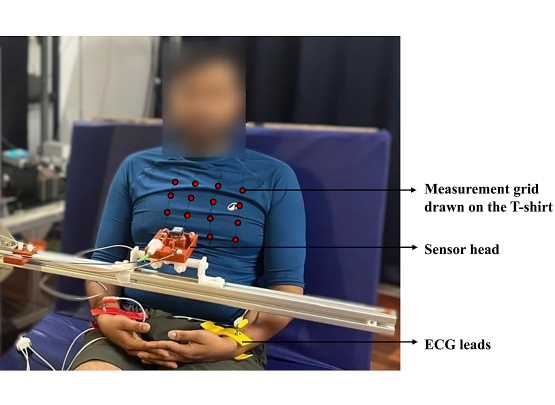}
		\caption{}
		\label{fig:subject}
	\end{subfigure}
	
	\caption{
		(a) Measurement grid marked with respect to anatomical landmarks with an intersensor spacing of 5 cm. (b) OPM sensor head used for recording MCG signals, with the measurement grid drawn on the subject’s T-shirt for sequential measurements.
	}
	\label{fig:grid_n_subject}
\end{figure*}

In this study, we present an unshielded MCG system based on OPMs operating in gradiometer mode which suppresses common-mode environmental noise for sequential acquisition across 16 thoracic positions. The gradiometric configuration suppresses common-mode environmental noise while preserving the spatial characteristics of cardiac magnetic fields. Comparing to multichannel systems, the proposed approach significantly reduces hardware complexity and cost, while enabling flexible spatial mapping.

	To improve the reliability of unshielded MCG measurements, we introduce a signal analysis framework for noise suppression and signal quality estimation (SQE) for automated beat quality assessment and rejection of contaminated cardiac cycles with noise. Using this framework, spatial magnetic field maps are reconstructed and analyzed during ventricular depolarization and repolarization. The reconstructed field maps are further compared with reference MCG datasets, both qualitatively and quantitatively, to evaluate morphological consistency and spatial agreement. By combining gradiometric OPM sensing, systematic signal quality enhancement, and spatial mapping, this work advances the feasibility of practical and cost-effective MCG systems for routine clinical and hospital-based cardiac diagnostics.
		
	\section{System Architecture and Methodology}
	  \subsection{Magnetocardiography}
MCG data were acquired from a healthy male volunteer (age: 35 years) using an in-house-developed rubidium vapor-cell OPM system operating in the Larmor regime \cite{patel2026magnetocardiography}. The subject was positioned in a semi-supine posture on a non-magnetic, non-conductive support structure, with the anterior thoracic surface maintained in close proximity to the sensor assembly, yielding a sensor-to-chest standoff distance of approximately 2–4 cm across measurement locations. A first-order axial gradiometer configuration was realized by operating two co-axially aligned magnetometers with an inter-sensor baseline of 3.5 cm; the differential output of the sensor pair was computed to suppress spatially correlated far-field ambient magnetic interference while preserving the near-field cardiac signal.
A standardized 16-point measurement grid was defined on the anterior thoracic surface using anatomical landmarks including the sternal notch, xiphoid process, and mid-clavicular lines as fiducial references, with adjacent grid points separated by a uniform 5 cm spacing to ensure adequate spatial sampling of the cardiac magnetic field distribution. Sequential single-position measurements were performed across all 16 thoracic locations, with each position recorded continuously for five minutes to enable ensemble-based signal averaging and noise characterization. To assess inter-trial repeatability and quantify measurement consistency, the full 16-position acquisition protocol was repeated twice in randomized position order, as illustrated in Figure \ref{fig:grid_n_subject}.
	
	 \begin{figure*}[t]
		\centering
		\includegraphics[
		width=0.85\textwidth,
		trim=0cm 4.5cm 0cm 4cm,
		clip
		]{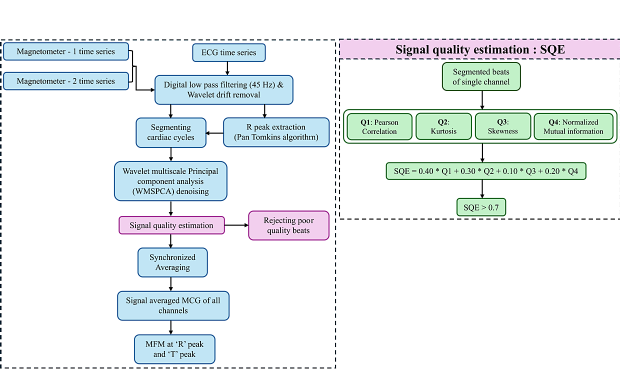}
		
		\caption{\textbf{Left block:} Flowchart of the proposed signal processing methodology. \textbf{Right block:} Signal quality estimation (SQE) framework.}
		\label{fig:flowchart}
	\end{figure*}
   \subsection{Signal Processing}
   MCG signals recorded in ambient environments are contaminated by various sources of environmental and physiological noise. The gradiometric configuration can remove common-mode noise present in both magnetometers; however, residual noise components remain and require additional signal processing. The raw MCG signals were recorded at a sample rate of 200 Hz and subsequently filtered using a digital finite impulse response low-pass filter with a cutoff frequency of 45 Hz to remove high-frequency non-cardiac noise present in the signals. The simultaneously recorded ECG signals in Lead-I configuration were used to obtain the R-peak time instances using the Pan–Tompkins algorithm \cite{pan1985realtime}. Based on the detected R-peak time instances, the MCG time series was segmented into individual cardiac-cycle beats containing the PQRST waveform.
   
   The segmented beats were further denoised using wavelet multiscale principal component analysis (WMSPCA), a hybrid signal-denoising technique that combines the multiresolution analysis capability of the discrete wavelet transform with the statistical dimensionality-reduction power of principal component analysis \cite{bakshi1998multiscale}. In this approach, each beat was decomposed up to level 4 using the db4 wavelet, after which PCA was applied across the beats at each decomposition level. Dominant principal components corresponding to the cardiac signal were retained, while uncorrelated noise components were suppressed. The variance threshold was set to 0.85, and the maximum number of principal components was capped at 5. At each wavelet decomposition level, the number of retained components was determined using the criterion that resulted in fewer principal components.
   Despite denoising, some beats may still contain motion artifacts, baseline drift, or sudden signal discontinuities caused by subject movement during acquisition. Therefore, an SQE framework was employed to identify and reject corrupted cardiac cycles prior to ensemble averaging. The complete methodology, including the SQE framework, is illustrated in Figure \ref{fig:flowchart}.
   
   \subsection{Signal Quality Estimation}
     \subsubsection{Pearson Correlation}
     The Pearson correlation coefficient between each individual beat and the
     averaged beat template was used to identify unrelated, noisy, or misaligned
     beats:
     \begin{equation}
     	\rho_i =
     	\frac{\operatorname{cov}\left(B_i, B_{\mathrm{ref}}\right)}
     	{\sigma_{B_i}\sigma_{B_{\mathrm{ref}}}},
     	\label{eq:pearson_correlation}
     \end{equation}
     where $B_i$ denotes the $i$th individual beat and $B_{\mathrm{ref}}$ denotes the
     reference beat template obtained by averaging across all beats. Negative
     correlation values were clipped to zero, yielding a bounded similarity score
     between 0 and 1,
     \begin{equation}
     	\rho_i^{+} = \max\left(0,\rho_i\right).
     	\label{eq:clipped_correlation}
     \end{equation}
     
     \subsubsection{Kurtosis}
   
     Kurtosis was used to identify prominent peaks in the beat waveform. In cardiac
     beats, the QRS complex produces a sharply peaked waveform; therefore, higher
     kurtosis values indicate the presence of pronounced QRS activity in the signal.
     For the $i$th beat, the excess kurtosis was calculated as
     \begin{equation}
     	\kappa_i =
     	\mathbb{E}\left[
     	\left(
     	\frac{B_i - \mu_{B_i}}{\sigma_{B_i}}
     	\right)^4
     	\right] - 3,
     	\label{eq:kurtosis}
     \end{equation}
     where $B_i$ is the individual beat, $\mu_{B_i}$ is its mean value, and
     $\sigma_{B_i}$ is its standard deviation.
     
     A normalized kurtosis score was then defined by comparing the kurtosis of the
     absolute beat waveform with that of the averaged reference beat:
     \begin{equation}
     	K_i =
     	\frac{\kappa\left(|B_i|\right)}
     	{\kappa\left(|B_{\mathrm{ref}}|\right)},
     	\label{eq:kurtosis_score}
     \end{equation}
     where $B_{\mathrm{ref}}$ is the averaged reference beat obtained by averaging
     across all beats. Negative kurtosis scores were clipped to zero, and scores
     greater than unity were clipped to one, yielding the final bounded kurtosis
     score
     \begin{equation}
     	K_i^{+} =
     	\min\left[1,\max\left(0,K_i\right)\right].
     	\label{eq:clipped_kurtosis_score}
     \end{equation}
     
     \subsubsection{Skewness}
    
     Skewness was used to quantify the asymmetry of the beat-amplitude distribution.
     Higher skewness values indicate the presence of asymmetric, QRS-complex-like
     features in the cardiac beat waveform. For the $i$th beat, the skewness was
     calculated as
     \begin{equation}
     	\gamma_i =
     	\mathbb{E}\left[
     	\left(
     	\frac{B_i - \mu_{B_i}}{\sigma_{B_i}}
     	\right)^3
     	\right],
     	\label{eq:skewness}
     \end{equation}
     where $B_i$ is the individual beat, $\mu_{B_i}$ is its mean value, and
     $\sigma_{B_i}$ is its standard deviation.

     A normalized skewness score was then defined by comparing the skewness of the
     absolute beat waveform with that of the averaged reference beat:
     \begin{equation}
     	S_i =
     	\frac{\gamma\left(|B_i|\right)}
     	{\gamma\left(|B_{\mathrm{ref}}|\right)},
     	\label{eq:skewness_score}
     \end{equation}

     where $B_{\mathrm{ref}}$ is the averaged reference beat obtained by averaging
     across all beats. Similar to the kurtosis score, beats with negative skewness
     scores or skewness scores greater than 2.0 were rejected by assigning their
     scores to zero. The remaining scores were scaled to obtain a bounded final
     skewness score:
     \begin{equation}
     	S_i^{+} =
     	\begin{cases}
     		S_i/2, & 0 \leq S_i \leq 2, \\
     		0, & S_i < 0 \ \mathrm{or}\ S_i > 2.
     	\end{cases}
     	\label{eq:clipped_skewness_score}
     \end{equation}
    
    \subsubsection{Normalized Mutual Information}
    Normalized mutual information (NMI) was used as a nonlinear metric to quantify
    the similarity between each individual beat and the averaged beat template. The
    NMI between the $i$th beat, $B_i$, and the reference beat, $B_{\mathrm{ref}}$,
    was defined as
    \begin{equation}
    	\mathrm{NMI}\left(B_i,B_{\mathrm{ref}}\right)
    	=
    	\frac{I\left(B_i;B_{\mathrm{ref}}\right)}
    	{\sqrt{H\left(B_i\right)H\left(B_{\mathrm{ref}}\right)}} ,
    	\label{eq:nmi_compact}
    \end{equation}
    where $I\left(B_i;B_{\mathrm{ref}}\right)$ is the mutual information between
    $B_i$ and $B_{\mathrm{ref}}$, and $H(\cdot)$ denotes the Shannon entropy.
    
    The mutual information was calculated as
    \begin{equation}
    	I\left(B_i;B_{\mathrm{ref}}\right)
    	=
    	\sum_{m=1}^{K}
    	\sum_{n=1}^{K}
    	p_{B_iB_{\mathrm{ref}}}(m,n)
    	\log
    	\left[
    	\frac{
    		p_{B_iB_{\mathrm{ref}}}(m,n)
    	}{
    		p_{B_i}(m)p_{B_{\mathrm{ref}}}(n)
    	}
    	\right],
    	\label{eq:mutual_information}
    \end{equation}
    and the corresponding entropies were calculated as
    \begin{equation}
    	H\left(B_i\right)
    	=
    	-\sum_{m=1}^{K}p_{B_i}(m)\log p_{B_i}(m),
    	\label{eq:entropy_bi}
    \end{equation}
    \begin{equation}
    	H\left(B_{\mathrm{ref}}\right)
    	=
    	-\sum_{n=1}^{K}p_{B_{\mathrm{ref}}}(n)
    	\log p_{B_{\mathrm{ref}}}(n).
    	\label{eq:entropy_bref}
    \end{equation}
    Here, $p_{B_i}(m)$ is the marginal probability distribution of the magnetic
    field values in $B_i$ corresponding to the $m$th histogram bin, and
    $p_{B_{\mathrm{ref}}}(n)$ is the marginal probability distribution of the
    magnetic field values in $B_{\mathrm{ref}}$ corresponding to the $n$th histogram
    bin. The term $p_{B_iB_{\mathrm{ref}}}(m,n)$ denotes the joint probability
    distribution between $B_i$ and $B_{\mathrm{ref}}$, representing the probability
    that a pair of corresponding field values simultaneously fall into bins $m$ and
    $n$, respectively. Both the individual beat and the averaged reference beat were
    discretized into equal-width histograms with $K=20$ bins. The resulting NMI
    score was bounded within the range $0$ to $1$.
    
    The final SQE score was calculated as a weighted
    linear combination of the extracted feature scores:
    \begin{equation}
    	\mathrm{SQE}_i =
    	0.4\rho_i^{+}
    	+ 0.3K_i^{+}
    	+ 0.1S_i^{+}
    	+ 0.2\mathrm{NMI}_i ,
    	\label{eq:sqe_score}
    \end{equation}
    where $\rho_i^{+}$, $K_i^{+}$, $S_i^{+}$, and $\mathrm{NMI}_i$ denote the
    clipped Pearson-correlation score, kurtosis score, skewness score, and
    normalized mutual-information score, respectively. Beats with an SQE score
    greater than $0.7$ were considered acceptable for further processing, whereas
    the remaining beats were excluded from subsequent analysis.
    
    The data are presented as mean $\pm$ standard deviation. All analyses and
    statistical tests, including Student's $t$-test and the Wilcoxon signed-rank
    test, were performed using the open-source Python 3.10.19 environment.

    \subsection{Magnetic Field Map (MFM)}
    The magnetic field recordings acquired at all measurement locations were used to reconstruct MFMs at specific cardiac time instants, such as the R-peak and T-peak. For each selected time point, the measured magnetic field values were spatially interpolated using cubic spline interpolation to estimate the field distribution between measurement locations. Iso-field contour maps were then generated by assigning separate colour scales to positive and negative magnetic field regions, producing a bipolar field pattern corresponding to the cardiac dipole \cite{fenici2005clinical, lim2009usefulness}. Similarly, the pseudo-current density vector was estimated by the gradient of the measured magnetic fields and overlaid on the MFM to indicate the direction of cardiac current flow during different phases of the cardiac cycle \cite{kandori1996reconstruction, haberkorn2006pseudo}. The spatial distribution of the MFM and the associated current-flow direction reflect the underlying cardiac conduction behaviour. Electrical inhomogeneities arising from cardiac diseases are reflected in the morphology, orientation, and spatial distribution of the magnetic field poles and current vectors \cite{pena2020ninety, tao2025aienabled, zhao2018integrated}. The orientation of the magnetic field pattern was characterized using the MFM angle, defined as the angle between the horizontal reference axis and the line connecting the positive and negative magnetic field poles \cite{lim2009usefulness}.
    
    \subsection{Signal Evaluation Metrics}
    The signal-to-noise ratio (SNR) was estimated by considering the QRS-complex
    interval as the signal regime, while the segment before the onset of the P wave
    was considered the noise regime, where no cardiac events occur. The ratio of the
    RMS values of the signals in these two regimes was converted to decibels for
    proper scaling:
    \begin{equation}
    	\mathrm{SNR}_{\mathrm{dB}}
    	=
    	20\log
    	\left(
    	\frac{\mathrm{RMS}_{\mathrm{QRS}}}
    	{\mathrm{RMS}_{\mathrm{iso}}}
    	\right),
    	\label{eq:snr_db}
    \end{equation}
    where $\mathrm{RMS}_{\mathrm{QRS}}$ is the RMS value calculated over the
    QRS-complex interval, and $\mathrm{RMS}_{\mathrm{iso}}$ is the RMS value
    calculated over the isoelectric segment before the onset of the P wave.
    
    The common-mode rejection ratio (CMRR) was estimated by evaluating the
    suppression of common environmental magnetic noise in the gradiometric
    configuration. The common-mode component was approximated using the summed output
    of the two magnetometers, while the differential output represented the residual
    signal after gradient subtraction. The CMRR was calculated as the ratio of the
    standard deviation of the common-mode signal to that of the differential signal:
    \begin{equation}
    	\mathrm{CMRR}_{\mathrm{dB}}
    	=
    	20\log
    	\left(
    	\frac{\sigma_{M_1+M_2}}
    	{\sigma_{M_1-M_2}}
    	\right),
    	\label{eq:cmrr_std}
    \end{equation}
    where $\sigma_{M_1+M_2}$ is the standard deviation of the summed magnetometer
    signals, and $\sigma_{M_1-M_2}$ is the standard deviation of the differential
    magnetometer signal.
    
    The frequency-dependent CMRR was computed by transforming the magnetic-field
    time-series signals into the frequency domain using power spectral density
    (PSD) analysis:
    \begin{equation}
    	\mathrm{CMRR}(f)_{\mathrm{dB}}
    	=
    	10\log
    	\left(
    	\frac{\mathrm{PSD}_{M_1+M_2}(f)}
    	{\mathrm{PSD}_{M_1-M_2}(f)}
    	\right),
    	\label{eq:cmrr_psd}
    \end{equation}
    where $M_1$ and $M_2$ are the time series of the two magnetometer signals. 
    
    The frequency-domain linear relationship between two magnetometer signals, 
    $M_1(t)$ and $M_2(t)$ was quantified using the magnitude-squared coherence:
    
    \begin{equation}
    	\gamma_{M_1 M_2}^2(f) = \frac{|G_{M_1 M_2}(f)|^2}{G_{M_1 M_1}(f) G_{M_2 M_2}(f)}
    \end{equation}
    
    where $G_{M_1 M_2}(f)$ is the cross-spectral density, and $G_{M_1 M_1}(f)$ 
    and $G_{M_2 M_2}(f)$ are the corresponding auto-spectral densities. 
    Coherence value ranges from 0 to 1, with higher values indicating 
    stronger linear dependence at a given frequency.
   
     \begin{figure*}[!t]
    	\centering
    	
    	\begin{subfigure}[t]{0.48\textwidth}
    		\centering
    		\includegraphics[width=\linewidth]{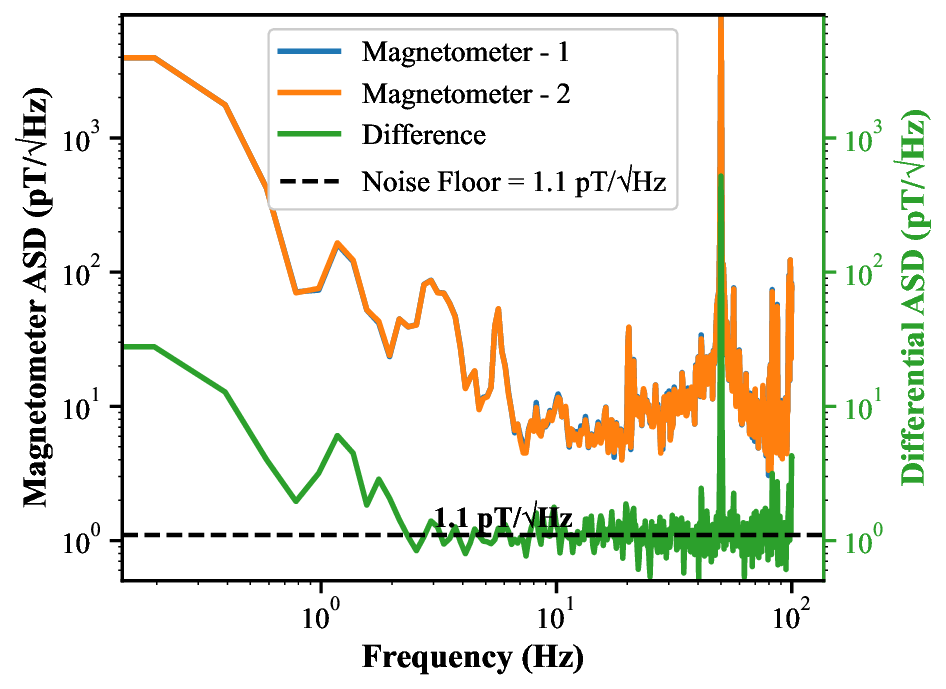}
    		\caption{}
    		\label{fig:sensitivity}
    	\end{subfigure}
    	\hfill
    	\begin{subfigure}[t]{0.48\textwidth}
    		\centering
    		\includegraphics[width=\linewidth]{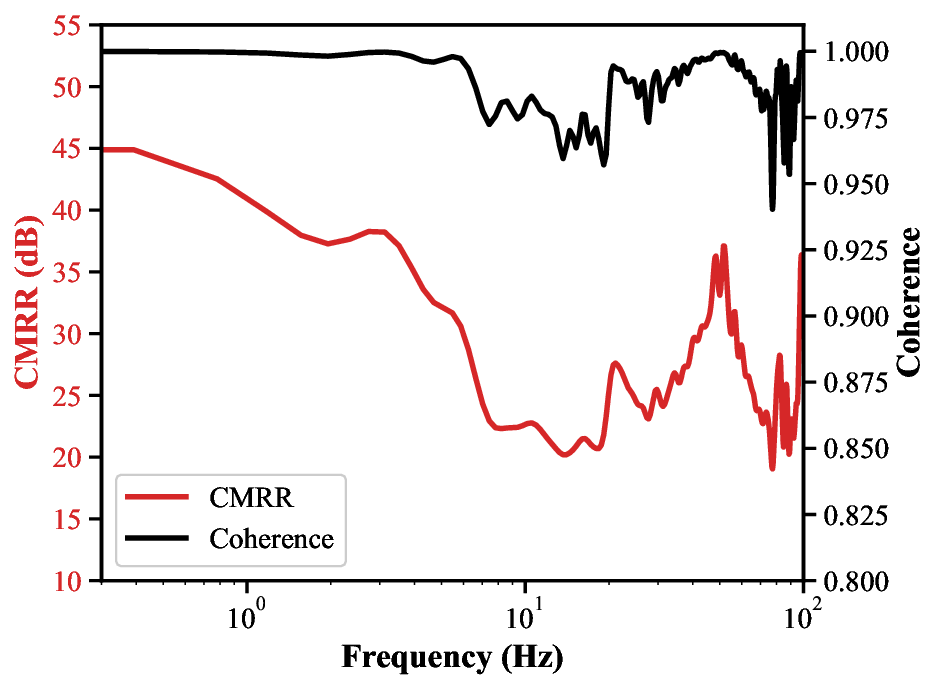}
    		\caption{}
    		\label{fig:cmrr}
    	\end{subfigure}

    	\caption{a. Amplitude spectral densities (ASDs) of Magnetometer 1 and Magnetometer 2 across different frequencies, along with the ASD of the gradiometric differential signal. b. Common-mode rejection ratio (CMRR) and coherence between the two magnetometers plotted as a function of frequency.}
    	\label{fig:sense_cmrr}
    \end{figure*}
    
    \begin{figure*}[!t]
    	\centering
    	\includegraphics[
    	width=0.85\textwidth]{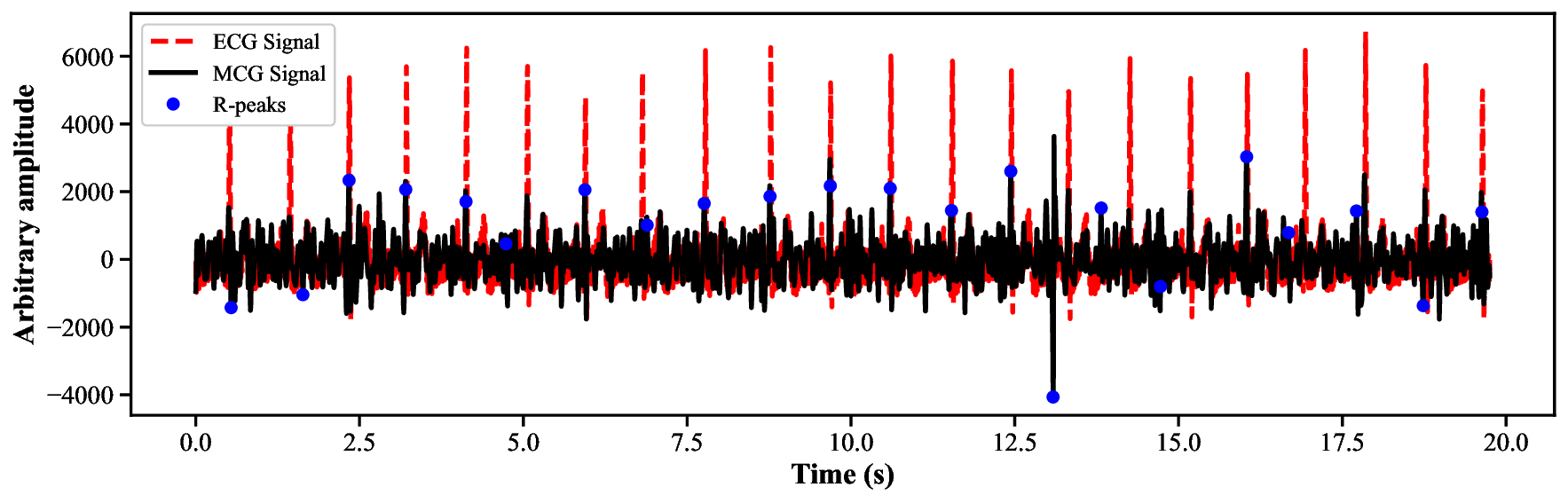}
    	
    	\caption{Measured MCG signals plotted along with simultaneously recorded ECG signals. The R-peak instances detected in the MCG signals (black traces) are marked with blue circles and show agreement with the ECG R peaks (red dashed lines).}
    	\label{fig:mcg_plot}
    \end{figure*}
    
	\section{Results}
	
	The magnetometer signals acquired in the ambient environment were significantly
	contaminated by external electromagnetic interference and environmental noise
	artifacts. The dominant frequency components of these artifacts are illustrated
	in Figure \ref{fig:sensitivity}. By applying gradiometric subtraction, the common-mode noise present in
	both magnetometers was effectively suppressed, resulting in a substantial
	reduction in the noise floor to approximately
	1.1~pT$/\sqrt{\mathrm{Hz}}$, corresponding to a gradiometer sensitivity of
	314~fT/cm$/\sqrt{\mathrm{Hz}}$.
	
	\begin{figure*}[!t]
		\centering
		
		\begin{subfigure}[t]{0.32\textwidth}
			\centering
			\includegraphics[width=\linewidth]{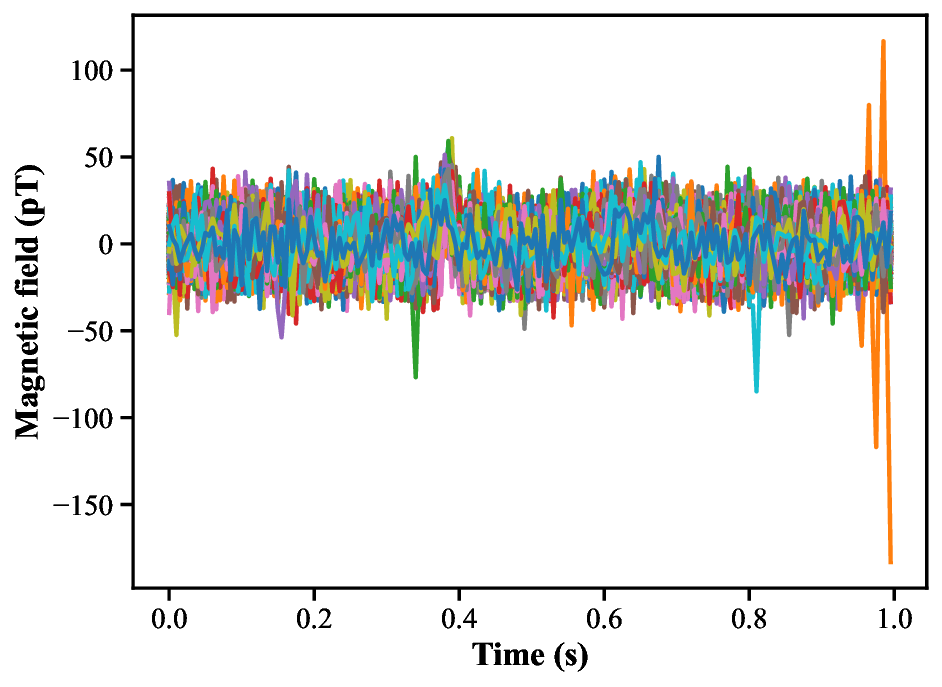}
			\caption{}
			\label{fig:mcg_signals_a}
		\end{subfigure}
		\hfill
		\begin{subfigure}[t]{0.32\textwidth}
			\centering
			\includegraphics[width=\linewidth]{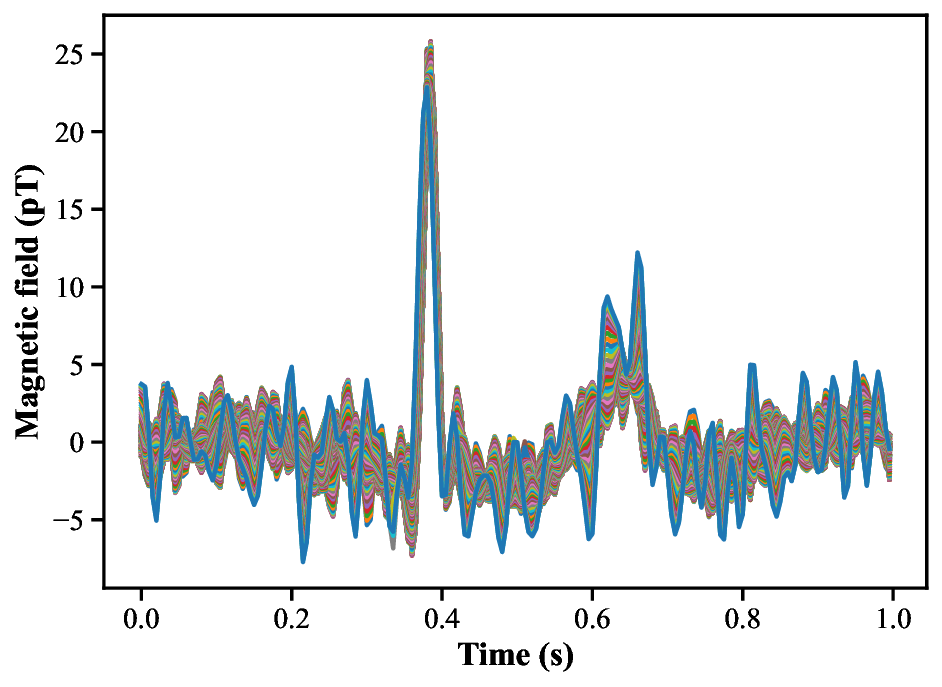}
			\caption{}
			\label{fig:mcg_signals_b}
		\end{subfigure}
		\hfill
		\begin{subfigure}[t]{0.32\textwidth}
			\centering
			\includegraphics[width=\linewidth]{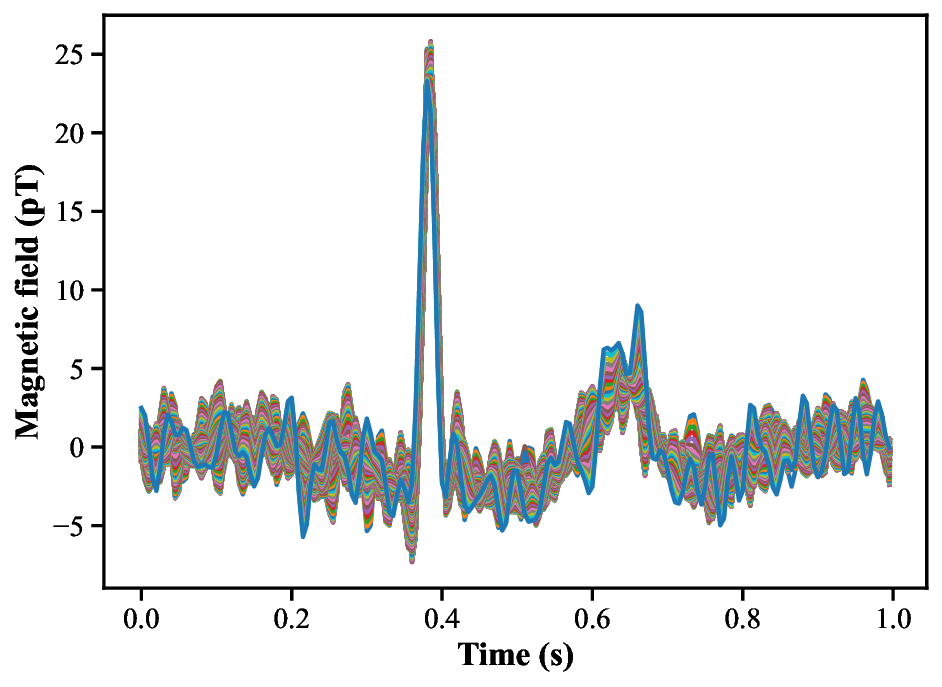}
			\caption{}
			\label{fig:mcg_signals_c}
		\end{subfigure}
		
		\vspace{0.4cm}
		
		\begin{subfigure}[t]{0.32\textwidth}
			\centering
			\includegraphics[width=\linewidth]{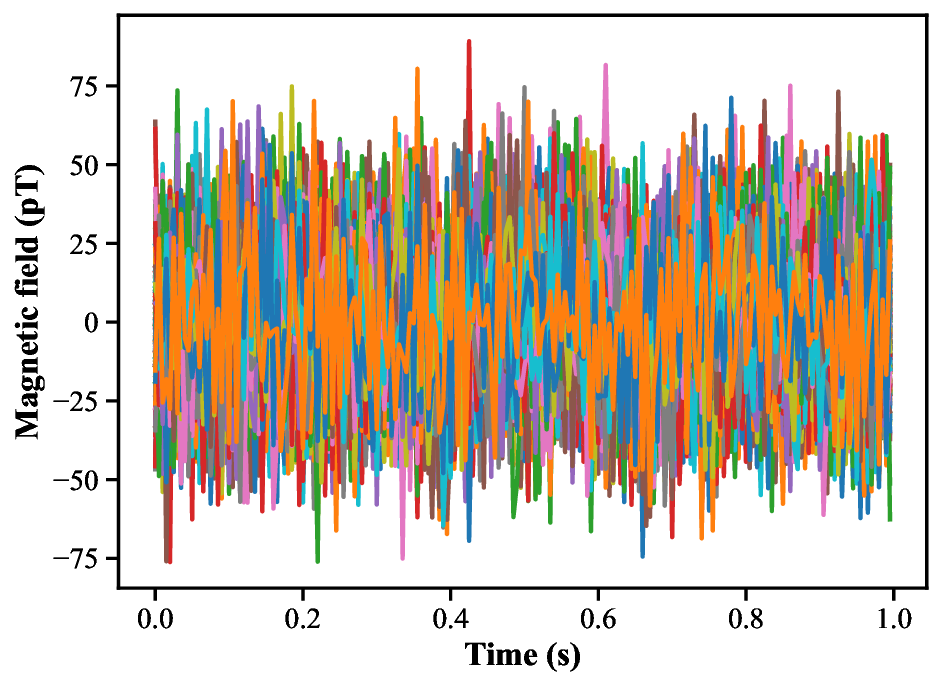}
			\caption{}
			\label{fig:mcg_signals_d}
		\end{subfigure}
		\hfill
		\begin{subfigure}[t]{0.32\textwidth}
			\centering
			\includegraphics[width=\linewidth]{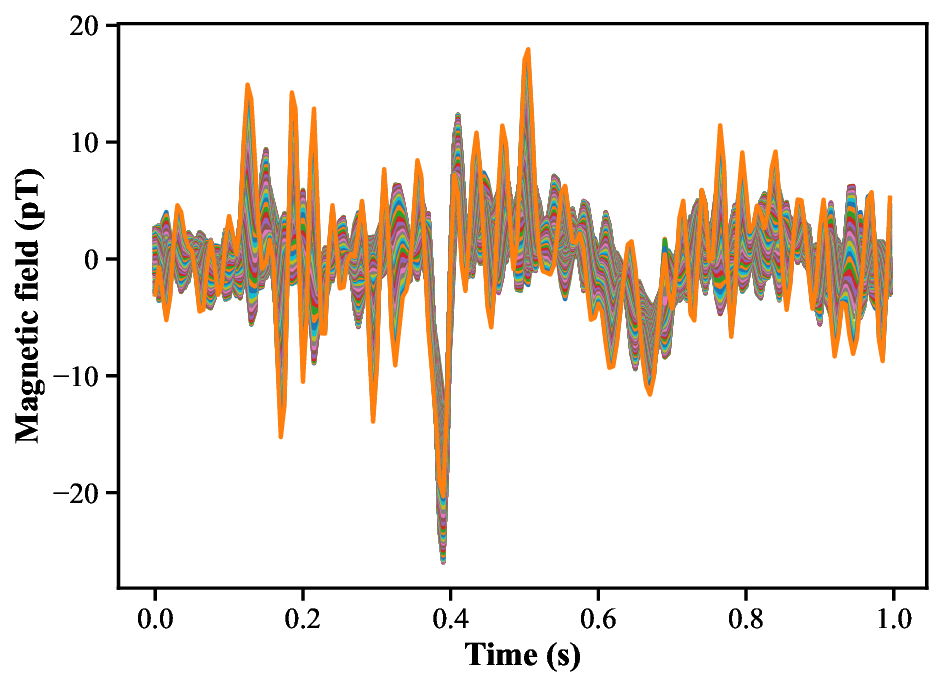}
			\caption{}
			\label{fig:mcg_signals_e}
		\end{subfigure}
		\hfill
		\begin{subfigure}[t]{0.32\textwidth}
			\centering
			\includegraphics[width=\linewidth]{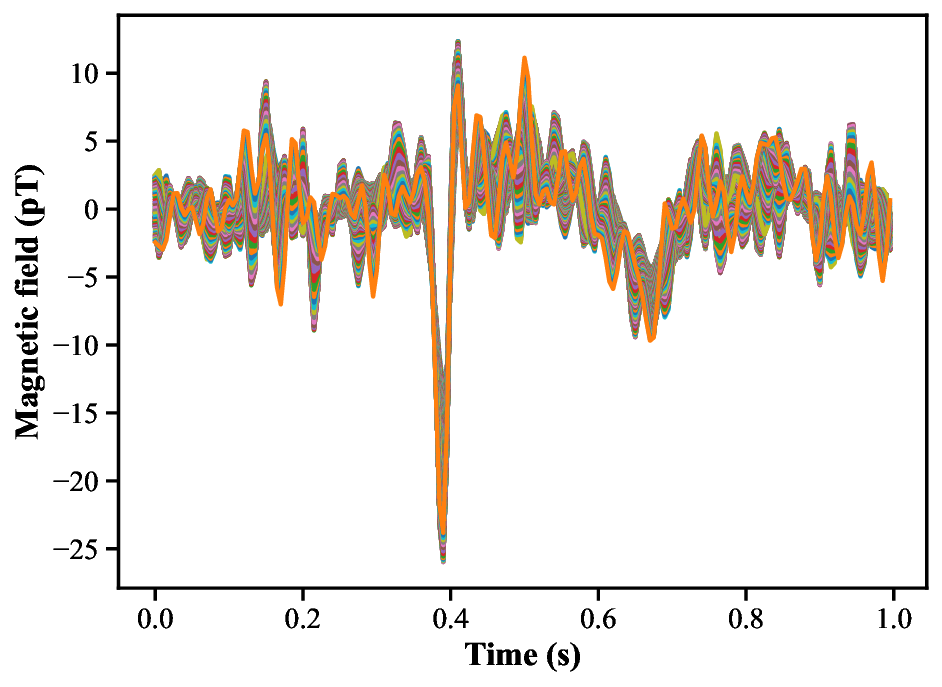}
			\caption{}
			\label{fig:mcg_signals_f}
		\end{subfigure}
		
		\caption{
			Raw MCG signals processed using the proposed signal processing framework to improve the SNR of the averaged beat. (a) Raw MCG signal segmented using ECG R peaks. (b) Signal after the proposed method of filtering. (c) Signal after removal of noisy beats using SQE for a particular measurement position. Similarly, (d), (e), and (f) represent the corresponding raw, filtered, and SQE-processed signals for another measurement position exhibiting different waveform polarity and morphology compared to the previous position. 
		}
		\label{fig:mcg_signals}
	\end{figure*}

	The common noise components between the two magnetometers were further quantified
	using coherence analysis and CMRR, as shown in Figure \ref{fig:cmrr}. The coherence values
	remained close to unity below approximately 6~Hz, indicating the presence of
	highly correlated low-frequency environmental noise in both channels. Beyond this
	range, the coherence exhibited moderate fluctuations with localized reductions,
	particularly in the higher-frequency region above 50~Hz. The lowest coherence
	values were observed around the 70--90~Hz band, suggesting the presence of
	partially uncorrelated high-frequency noise components.
	
	The time-domain CMRR was approximately 31~dB. In addition, the frequency-dependent
	CMRR is presented in Figure \ref{fig:cmrr}. Higher CMRR values were obtained in the low-frequency
	regime, while the rejection capability gradually decreased at higher frequencies.
	Nevertheless, the CMRR remained above approximately 20~dB within most of the
	cardiac signal bandwidth. These results indicate that gradiometric subtraction is
	highly effective in suppressing low-frequency common-mode interference, whereas
	the rejection performance decreases for uncorrelated high-frequency noise
	components. Since the dominant spectral components of cardiac magnetic signals
	lie within the 0.5--40~Hz range, the achieved CMRR is adequate for effective
	background-noise suppression during MCG measurements.
	
	The low-pass-filtered MCG signal acquired at sensor position [3,2] is shown in
	Figure \ref{fig:mcg_plot}. Since this location is situated close to the cardiac region, it exhibited
	a comparatively higher SNR, enabling clear visualization of the R-peaks. However,
	the ECG-derived heart rate showed stable behavior
	$(69.23 \pm 3.71)$~bpm, whereas the raw MCG-derived heart rate exhibited
	substantially higher variability $(70.49 \pm 11.29)$~bpm, indicating the
	presence of intermittent noise and motion artifacts during measurement, which
	resulted in false peak detections in the digitally filtered MCG signals. A paired
	Student's $t$-test revealed a statistically significant difference between the
	ECG- and MCG-derived heart-rate estimates $(p = 0.00043)$, demonstrating the
	necessity of signal preprocessing and artifact rejection prior to further
	analysis.
	
	\begin{figure*}[t]
		\centering
		
		\begin{subfigure}[t]{0.48\textwidth}
			\centering
			\includegraphics[width=\linewidth]{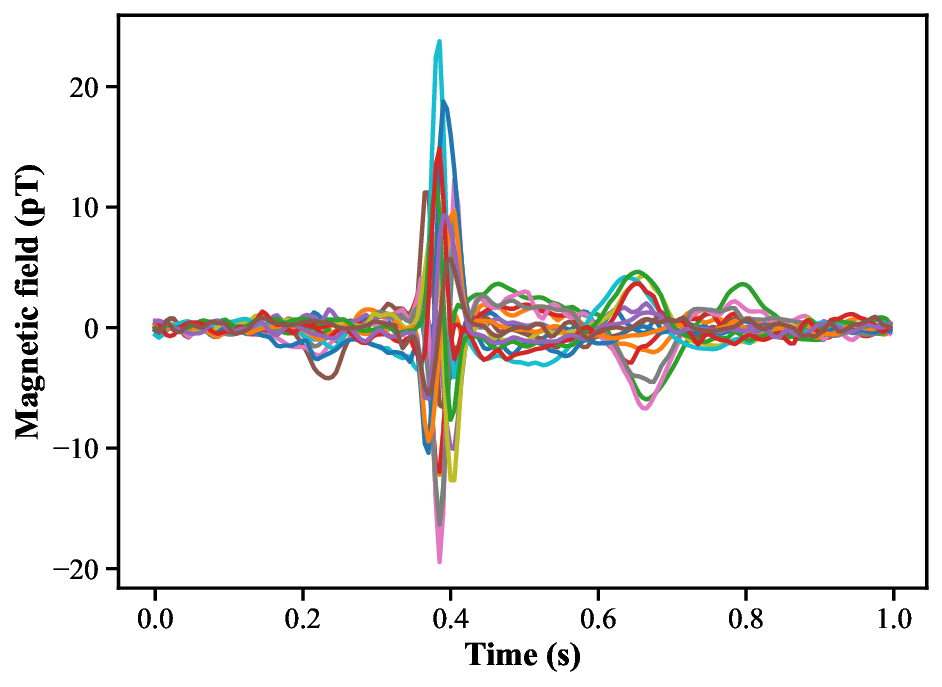}
			\caption{}
			\label{fig:signal_averaged_beat_a}
		\end{subfigure}
		\hfill
		\begin{subfigure}[t]{0.42\textwidth}
			\centering
			\includegraphics[width=\linewidth]{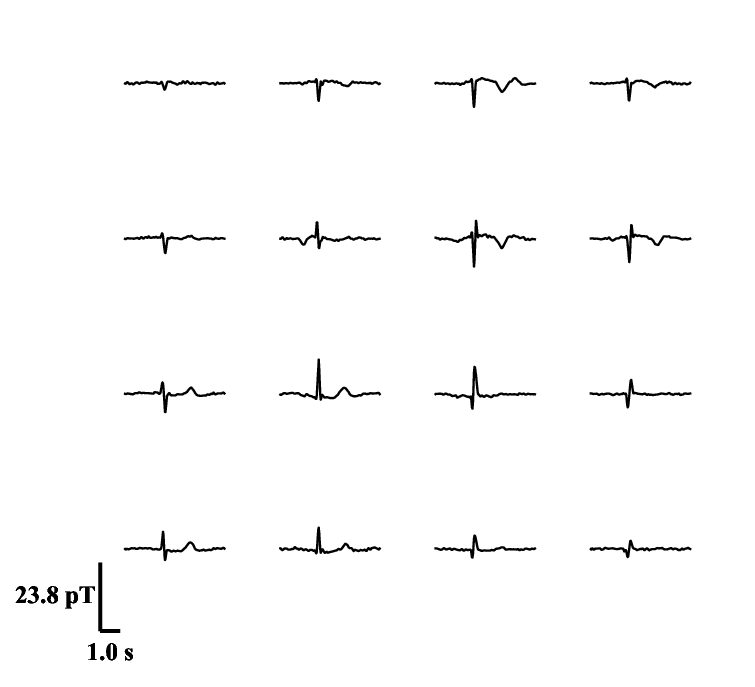}
			\caption{}
			\label{fig:signal_averaged_beat_b}
		\end{subfigure}
		
		\caption{
			(a) Signal-averaged beat obtained after the complete denoising process for all the sensor channels (Butterfly plot). (b) Spatial representation of the averaged beats with respect to sensor position, illustrating polarity reversal across spatial locations.
		}
		\label{fig:signal_averaged_beat}
	\end{figure*}
	
	The digitally filtered MCG time-series data were segmented into individual
	cardiac cycles based on ECG-derived R-peak time instances, since direct
	MCG-based R-peak detection was found to be susceptible to noise and false
	detections. Each segmented beat contained complete PQRST waveforms, as
	illustrated in the workflow presented in Figure \ref{fig:mcg_signals}. Initially, the segmented beat
	signals were heavily contaminated with noise, making the cardiac morphology
	difficult to identify. To improve signal quality, the segmented beats were
	denoised using the WMSPCA algorithm, which effectively suppressed uncorrelated
	noise components. Following denoising, the characteristic cardiac waveform
	morphology became clearly distinguishable. Subsequently, SQE-based beat analysis
	was applied to eliminate noisy beats while retaining only high-quality beats for
	ensemble averaging.
	
	Similarly, for position [2,4], the raw segmented beat signals exhibited poor
	visibility of cardiac waveform prior to denoising. After the proposed denoising
	and beat-selection procedure, the signal quality improved considerably, enabling
	clear identification of cardiac features with enhanced SNR. Figure \ref{fig:mcg_signals}
	present representative results obtained from sensor position [2,4]. Furthermore,
	the polarity reversal phenomenon observed between the two sensor positions
	became distinctly visible only after noise suppression, highlighting the spatial
	distribution of the cardiac magnetic field and the necessity of the proposed
	signal-denoising method.

	The denoised beats from all sensor locations were ensemble-averaged to further
	improve the SNR. The averaged waveforms were subsequently smoothed and
	mean-corrected to obtain the final representative cardiac magnetic signals. The
	resulting averaged waveforms across all sensor locations are illustrated as a
	butterfly plot in Figure \ref{fig:signal_averaged_beat_a}. A clear polarity-reversal pattern can be observed
	in Figure \ref{fig:signal_averaged_beat_b}, reflecting the dipolar nature of the cardiac magnetic field
	distribution. The spatial variation in waveform polarity across sensor locations
	demonstrates the ability of the proposed system to capture physiologically
	meaningful cardiac magnetic field patterns.
	
\begin{figure*}[t]
		\centering
		\includegraphics[width=0.75\textwidth]{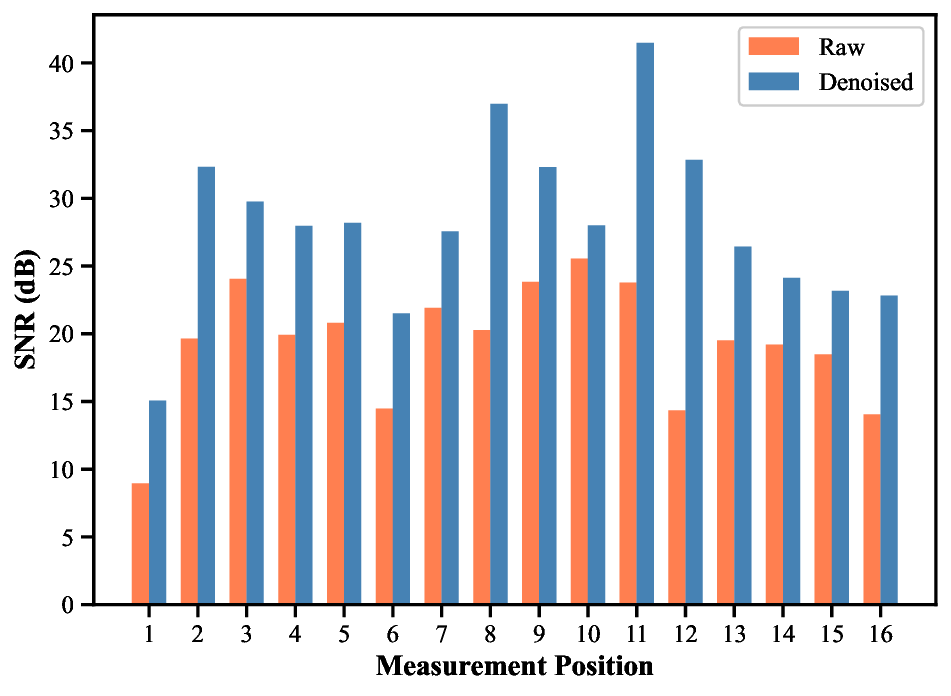}
		
		\caption{QRS SNR values computed for denoised and averaged signals at each measurement location, compared with the SNR of the raw averaged signals to demonstrate the improvement achieved after filtering.}
		\label{fig:QRS_SNR}
	\end{figure*}

	A substantial enhancement in signal quality was observed after denoising and
	beat-selection processing. The SNR improvement achieved using the proposed
	signal-processing framework across all sensor locations is presented in Figure \ref{fig:QRS_SNR}.
	Statistical analysis using a paired Student's $t$-test revealed that the
	denoised signals exhibited significantly higher SNR compared with the raw signals
	$(t(15) = 7.22, p < 0.001)$. The statistical significance of this improvement was
	further confirmed using the Wilcoxon signed-rank test $(p < 0.001)$. These
	results validate the effectiveness of the proposed processing methodology for
	improving the quality of unshielded MCG recordings.
	
	After signal-quality enhancement and averaging, the reconstructed MCG signals
	were used to generate MFMs at the R-peak and T-peak time instants. Since
	CAD-associated electrical inhomogeneities are predominantly reflected during
	ventricular depolarization and repolarization, MFMs were analyzed at these
	physiologically significant intervals. The obtained MFMs at the R-peak and
	T-peak generally exhibited similar orientation and spatial distribution patterns,
	consistent with expected cardiac electrophysiological behavior.
	
\begin{figure*}[!t]
		\centering
		
		\begin{subfigure}[t]{0.25\textwidth}
			\centering
			\includegraphics[width=\linewidth]{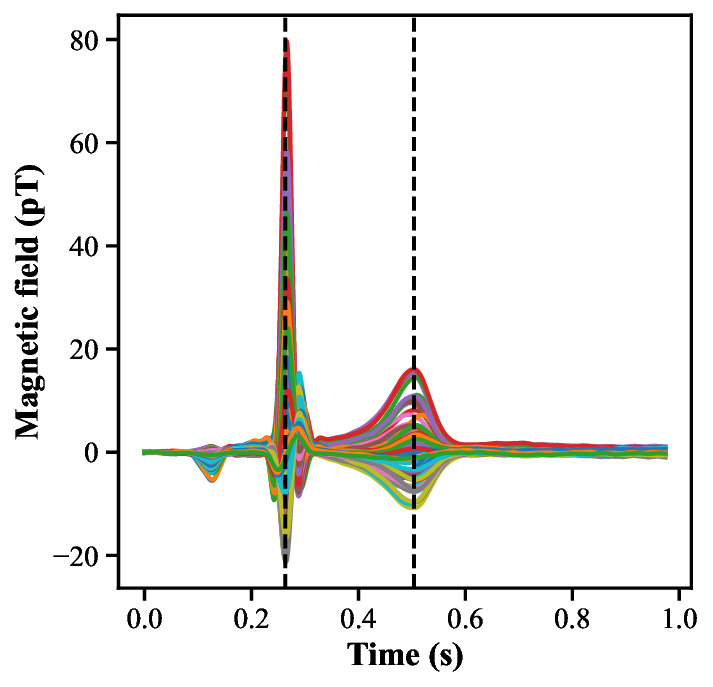}
			\caption{}
			\label{fig:butterfly_mfm_a}
		\end{subfigure}
		\hfill
		\begin{subfigure}[t]{0.30\textwidth}
			\centering
			\includegraphics[width=\linewidth]{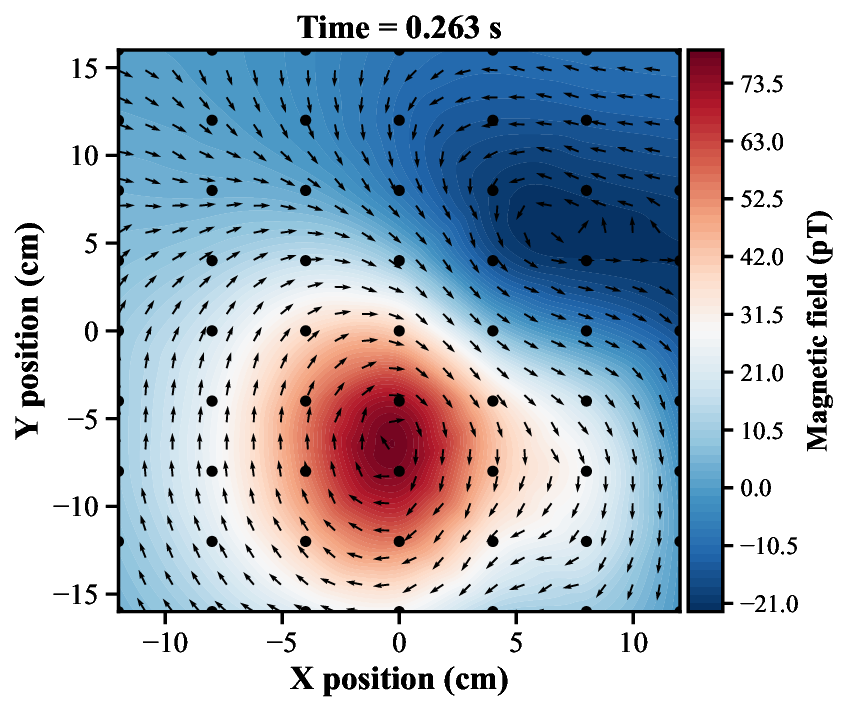}
			\caption{}
			\label{fig:butterfly_mfm_b}
		\end{subfigure}
		\hfill
		\begin{subfigure}[t]{0.30\textwidth}
			\centering
			\includegraphics[width=\linewidth]{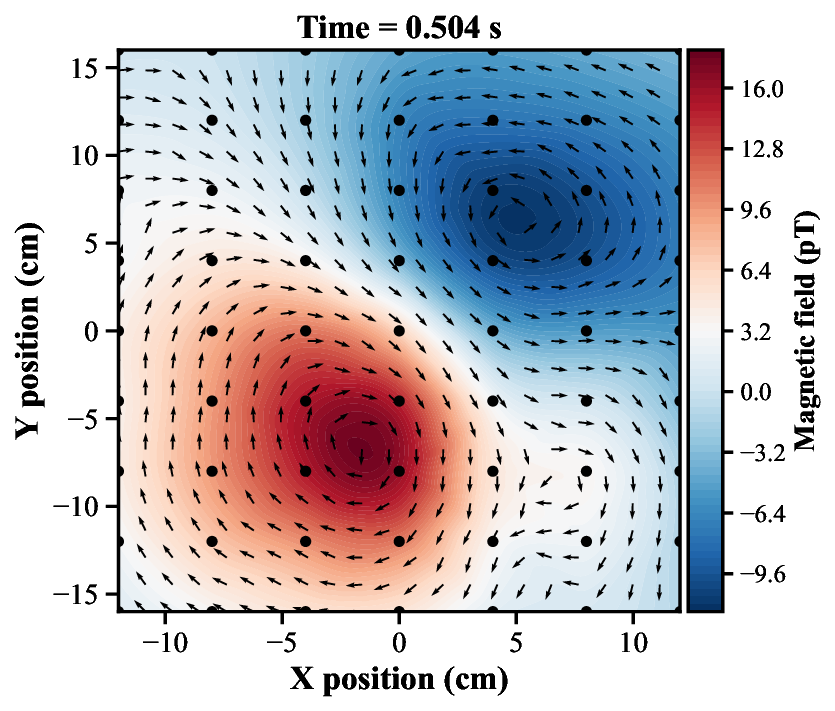}
			\caption{}
			\label{fig:butterfly_mfm_c}
		\end{subfigure}
		
		\vspace{0.4cm}
		
		\begin{subfigure}[t]{0.25\textwidth}
			\centering
			\includegraphics[width=\linewidth]{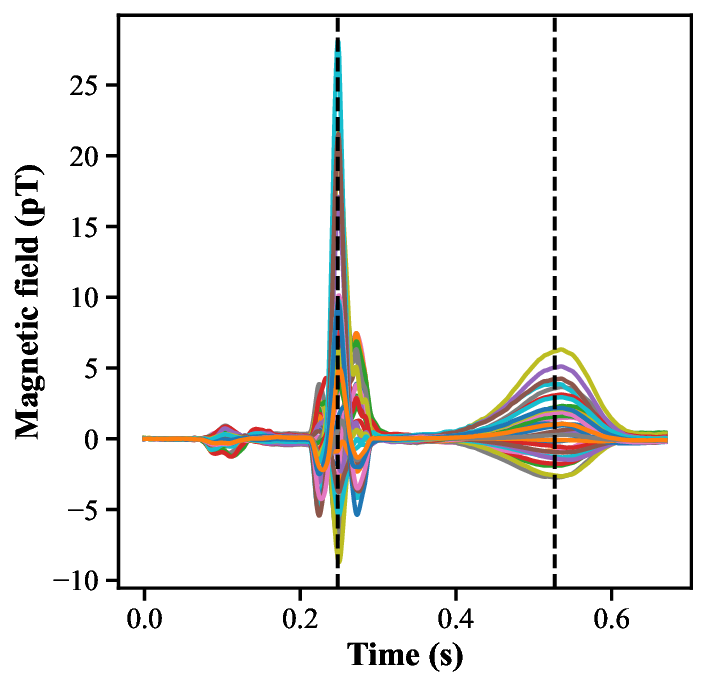}
			\caption{}
			\label{fig:butterfly_mfm_d}
		\end{subfigure}
		\hfill
		\begin{subfigure}[t]{0.30\textwidth}
			\centering
			\includegraphics[width=\linewidth]{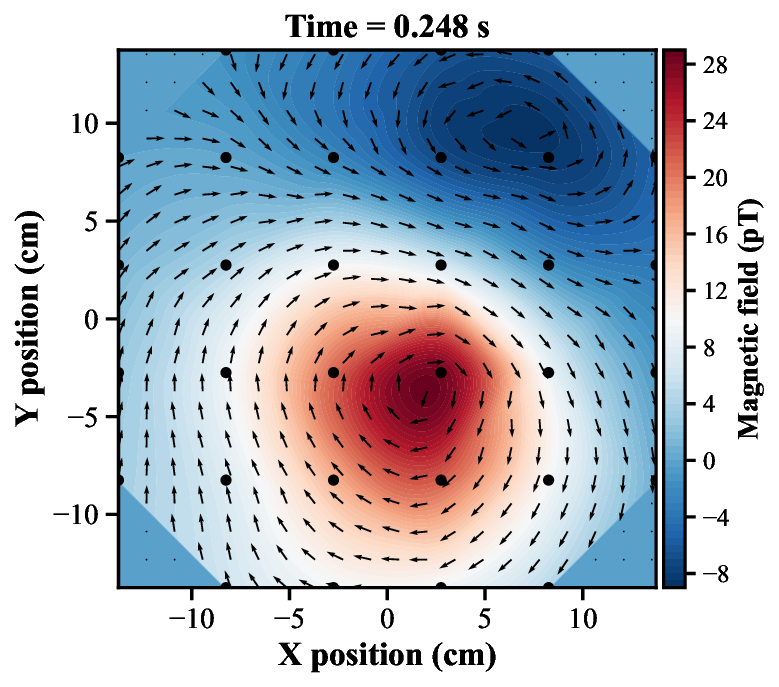}
			\caption{}
			\label{fig:butterfly_mfm_e}
		\end{subfigure}
		\hfill
		\begin{subfigure}[t]{0.30\textwidth}
			\centering
			\includegraphics[width=\linewidth]{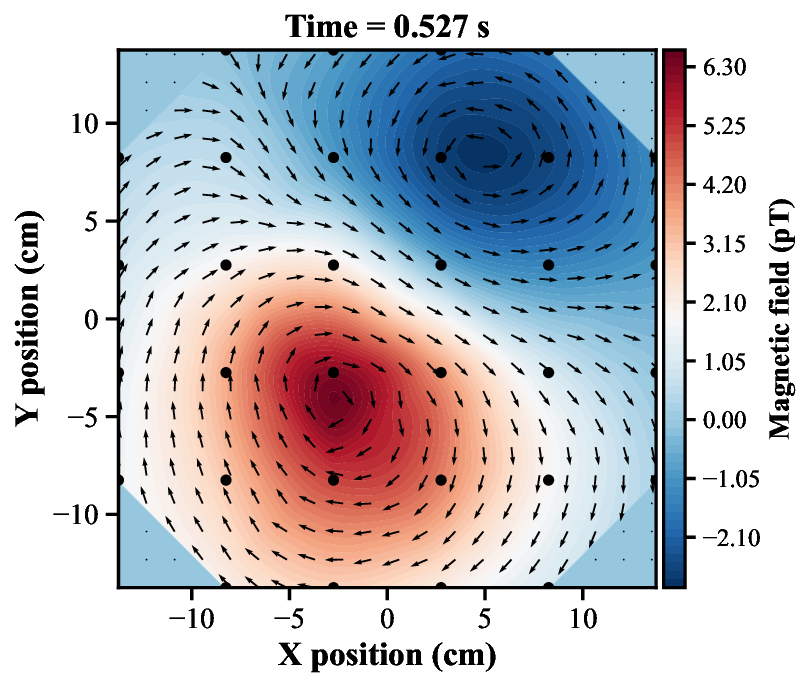}
			\caption{}
			\label{fig:butterfly_mfm_f}
		\end{subfigure}
		
		\vspace{0.4cm}
		
		\begin{subfigure}[t]{0.25\textwidth}
			\centering
			\includegraphics[width=\linewidth]{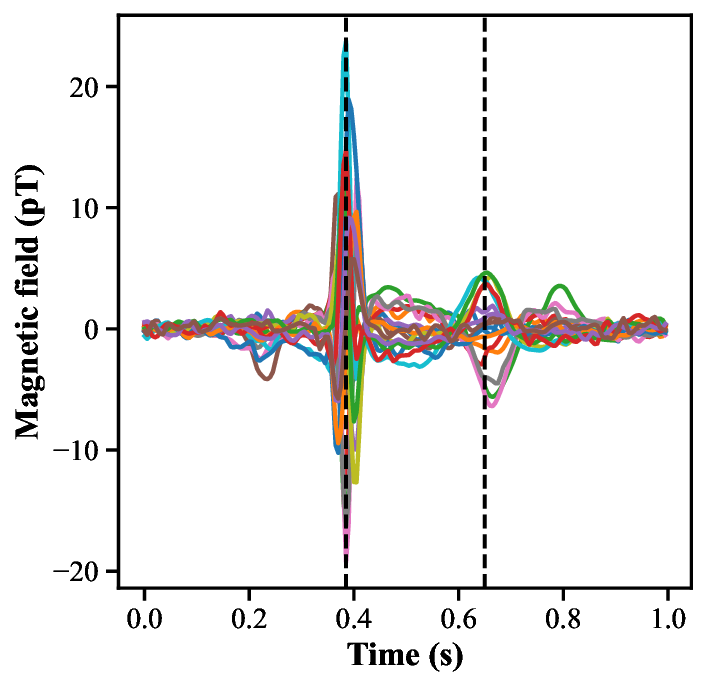}
			\caption{}
			\label{fig:butterfly_mfm_g}
		\end{subfigure}
		\hfill
		\begin{subfigure}[t]{0.30\textwidth}
			\centering
			\includegraphics[width=\linewidth]{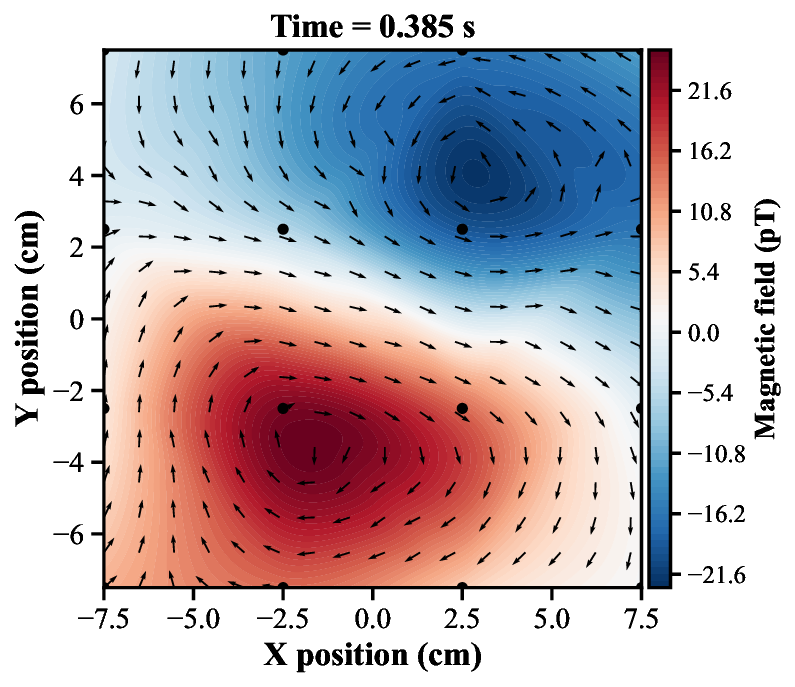}
			\caption{}
			\label{fig:butterfly_mfm_h}
		\end{subfigure}
		\hfill
		\begin{subfigure}[t]{0.30\textwidth}
			\centering
			\includegraphics[width=\linewidth]{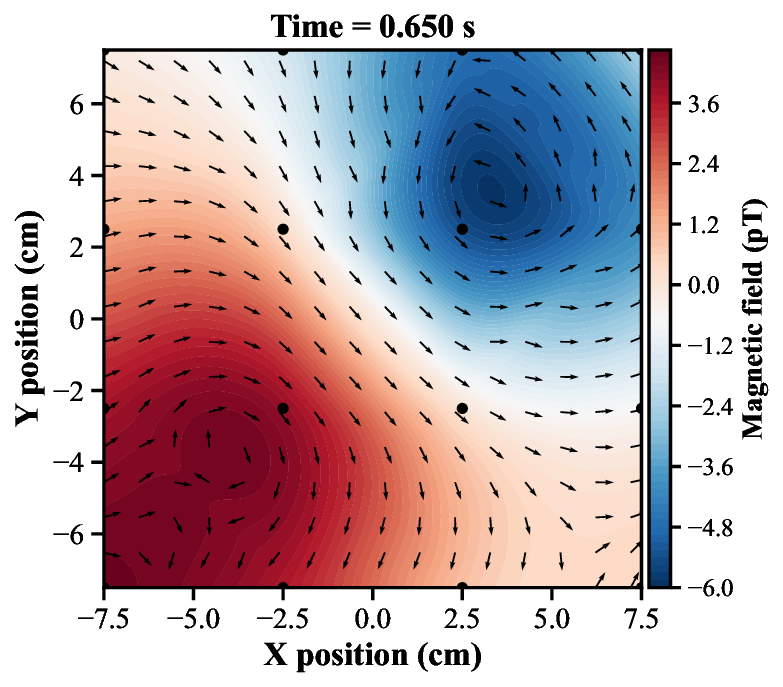}
			\caption{}
			\label{fig:butterfly_mfm_i}
		\end{subfigure}
		
		\caption{Comparison of MFMs with current density arrows indicating the direction of the cardiac current flow obtained from the measured MCG signals with reference MCG datasets. (a), (b), and (c) represent the butterfly plot of averaged MCG signals from 63 sensor channels, along with the MFMs at the R-peak and T-peak time instants, respectively. Similarly, (d), (e), and (f) represent the averaged MCG signals from 32 channels and the corresponding MFMs at the R-peak and T-peak instants. (g), (h), and (i) represent the measured MCG signals obtained using the 16-channel OPM sensor system with the proposed signal processing framework, along with the corresponding MFMs at the R-peak and T-peak instants. The black arrows overlaid on the contour map represent the direction of the estimated current density vector. The position of the sensors are marked by the black circles. The MFM angle parameter computed for each MFM is indicated within the respective plots. }
		\label{fig:butterfly_mfm}
	\end{figure*}

The measured unshielded MCG MFMs are shown in Figure \ref{fig:butterfly_mfm}, along with reference MFMs obtained from available online databases that were recorded inside magnetically shielded rooms using 63-channel signal-averaged MCG recordings acquired from a healthy subject using a SERF-OPM system with 4 cm inter-sensor spacing in a magnetically shielded environment \cite{su2024vector} and 32-channel  MCG recordings obtained from 20 subjects using a SERF-OPM system with 5 cm inter-sensor spacing under shielded conditions \cite{zhu2024mcgdataset}.  Among these subjects, only recordings corresponding to normal sinus rhythm and showing a clear dipole pattern were considered for the present study. The spatial and temporal distribution of the cardiac magnetic field was analyzed using MFMs. 
	
The MFM angles, representing the orientations
	of the magnetic field distributions, were computed for all datasets and annotated
	on the corresponding maps. In addition, the pseudo-current directions indicated
	by the black current arrows exhibited consistent directional behavior across all
	datasets. These observations strongly suggest that the proposed acquisition and
	signal-processing methodology preserves the essential electrophysiological
	characteristics of cardiac magnetic activity without significant information
	loss.
    \begin{table}[t]
    	\caption{\label{tab:mfm_feature_values}
    		MFM feature values.}
    	\centering
    	\footnotesize
    	\renewcommand{\arraystretch}{1.25}
    	\setlength{\tabcolsep}{5pt}
    	
    	\begin{tabular}{l c c}
    		\hline\hline
    		\textbf{Dataset} & \textbf{R peak} & \textbf{T peak} \\
    		& \textbf{(degrees)} & \textbf{(degrees)} \\
    		\hline
    		
    		63 sensor channels data &
    		$-60.30$ &
    		$-63.54$ \\
    		
    		32 sensor channels data &
    		$-61.42 \pm 7.31$ &
    		$-64.52 \pm 8.73$ \\
    		
    		Measured 16 channels data &
    		$-58.20$ &
    		$-45.79$ \\
    		
    		\hline\hline
    	\end{tabular}
    \end{table}
	The quantitative MFM angle measurements obtained at the R-peak and T-peak time
	instants are summarized in Table \ref{tab:mfm_feature_values}. The measured angles for the proposed
	16-channel unshielded MCG system were found to lie within the range of values
	reported for the reference shielded MCG databases. In particular, the R-peak and
	T-peak MFM angles of the measured data showed only minor deviations compared
	with the 63-channel and 32-channel reference datasets. This agreement indicates
	that the proposed system is capable of reproducing clinically relevant cardiac
	magnetic field orientations despite the reduced channel count and operation in
	an unshielded environment.

	\section{Discussion}
	In the present study, the proposed signal-processing framework enabled the
	extraction of physiologically meaningful cardiac magnetic signals despite
	operation in an ambient environment with substantial electromagnetic
	interference. The achieved common-mode noise suppression highlights an important
	advantage of OPM-based gradiometry, where nearly identical sensor characteristics
	improve common-noise cancellation compared with conventional systems requiring
	multiple reference channels~\cite{yaga2025recording}. Although the first-order gradiometer
	effectively reduced common environmental interference, gradient magnetic noise
	components remain a limitation. Higher-order gradiometer configurations may
	further improve noise suppression, but they may also introduce signal attenuation
	and increased system complexity~\cite{drung2002electronic, lee2007sixtyfour, zhang2020portable}.
	
	Reliable beat selection is essential in practical MCG measurements because
	ensemble averaging of several hundred cardiac cycles is generally required to
	obtain clinically usable signals. Noisy beats can significantly degrade the
	reconstructed waveform and spatial magnetic field distribution. In the proposed
	SQE framework, the ensemble-averaged beat was used as the reference template
	instead of employing cross-beat normalization~\cite{jia2025novel}. The averaged beat,
	obtained through coherent summation of all tentatively detected cardiac cycles,
	represents the best achievable signal morphology for a given measurement session
	and sensor configuration. This approach provides a physically interpretable and
	batch-independent quality metric, where higher SQE scores indicate closer
	agreement with the expected cardiac waveform. Beats with SQE scores below $0.7$
	were considered unsuitable for ensemble averaging and were excluded from further
	analysis.
	
	Signal quality analysis of MCG signals has been explored by several
	researchers. However, these studies were performed in shielded environments with
	better SNR in the acquired signals; hence, these methods may underperform in
	ambient environments~\cite{lu2018realtime,jia2025novel,zhang2009quantitative}. The proposed SQE-based
	beat-selection and WMSPCA-denoising framework substantially improved signal
	quality and enabled consistent recovery of cardiac morphology across different
	sensor positions. The statistically significant SNR improvement further supports
	the effectiveness of the proposed processing methodology.
	
	The MFMs and  pseudo-current density vectors reconstructed at the R-peak and T-peak time instants demonstrated spatial distributions and orientations comparable to those of existing shielded multichannel MCG datasets. In particular, MFM angles obtained from the proposed unshielded system showed quantitative agreement with reference datasets acquired using 63-channel and 32-channel shielded systems. The successful identification of consistent MFM orientation and pseudo-current-flow direction indicates that the proposed methodology preserves essential electrophysiological
	information despite operation in an unshielded environment with a limited number of channels~\cite{haberkorn2006pseudo}. Furthermore, the reliable detection of
	T-wave-associated magnetic field distributions highlights the potential
	feasibility of using the proposed system for assessing ventricular repolarization
	abnormalities~\cite{tao2025aienabled}.
	
	Although recent advances in room-temperature magnetic sensors have enabled the
	acquisition of MCG signals in semi-shielded and completely unshielded
	environments, most existing studies have primarily focused on signal acquisition
	at one or two measurement locations, highlighting the need for robust
	signal-processing frameworks to improve signal quality and reliability for
	routine MCG applications~\cite{xiao2023movable, limes2020portable, patel2026magnetocardiography}. The proposed framework can
	potentially be extended to other magnetic sensing technologies with comparable
	sensitivity to enhance cardiac signal extraction under ambient measurement
	conditions.
	
	The present study has a few limitations. First, the measurements were
	performed on a single healthy subject; therefore, large-scale validation
	involving both healthy subjects and patients with cardiac abnormalities is
	required to establish the clinical reliability of the proposed methodology.
	Variations in sensor-to-body distance and subject motion during prolonged
	acquisition may also influence waveform amplitude and the spatial magnetic field
	distribution.
	
	The proposed signal-processing framework is primarily optimized for subjects
	with normal sinus rhythm, where R-peak-based averaging effectively improves
	signal quality. Consequently, the methodology may be extended to CAD subjects
	maintaining stable sinus rhythm. However, in subjects with cardiac arrhythmias,
	significant beat-to-beat morphological variations may occur, and conventional
	averaging approaches may suppress clinically relevant transient
	features~\cite{mantynen2018noninvasive}. Furthermore, in the current SQE implementation, the
	feature-weighting coefficients and decision threshold were empirically selected
	based on the obtained experimental results. Further large-scale analysis is
	required to determine the optimal weighting strategy and threshold selection
	across different cardiac conditions and measurement environments with varying
	sources of noise and interference.
	
	In the present study, MFM analysis was performed using sequential recordings
	acquired from 16 measurement locations with a spacing of 5~cm between adjacent
	measurement points. Although this sequential acquisition approach reduces system
	complexity and overall instrumentation cost, conventional clinical MCG systems
	typically employ more than 30 simultaneously acquired channels to achieve higher
	spatial resolution of the cardiac magnetic field distribution~\cite{pena2020ninety, kandori1996reconstruction}.
	Furthermore, sequential measurements may increase the total acquisition duration
	and may fail to capture transient cardiac events, such as premature ventricular
	contractions, occurring between recording intervals~\cite{aita2019noninvasive}.
	
	Future work will focus on clinical validation involving larger healthy and
	pathological cohorts, integration of motion-compensation techniques, adaptive
	beat-selection methods for non-stationary rhythms, and optimized multichannel
	gradiometric configurations for improved spatial resolution and noise
	suppression.
	
	\section{Conclusion}
	This study demonstrates the feasibility of unshielded MCG using a single sensor having two rubidium-based OPMs operating in a gradiometer configuration for sequential
	cardiac magnetic field acquisition. The proposed system enabled reliable
	detection of cardiac magnetic signals in ambient environments while reducing the
	complexity and cost associated with conventional shielded multichannel MCG
	systems. A signal-processing framework was developed to improve signal
	reliability by suppressing environmental interference and rejecting corrupted
	cardiac cycles prior to averaging. Using sequential measurements acquired across
	16 thoracic positions, spatial magnetic field maps and pseudo-current-density
	distributions were reconstructed during ventricular depolarization and
	repolarization, showing good agreement with reference shielded Spin-Exchange Relaxation-Free (SERF)-OPM MCG
	datasets. Overall, the proposed approach demonstrates that clinically meaningful
	cardiac magnetic field mapping can be achieved using a compact and
	cost-effective gradiometric OPM system operating under ambient conditions.

	\begin{acknowledgments}
		The authors acknowledge the financial support and funding received from the Qmet Tech Foundation, Department of Science and Technology (DST) under the National Quantum Mission (NQM) of Government of India, for conducting the experiments in this study. Additionally, the authors thank Central Manufacturing Technology Institute (CMTI), Bangalore, for their technical support in the development of the spectroscopic setup.
		
	\end{acknowledgments}
	
	\section*{Data Availability}
	
	The data that support the findings of this study are available from the corresponding author upon reasonable request.
	
	\section*{References}
	\vspace*{-0.8cm}
	%\bibliography{references}
	%aipnum4-2.bst 2019-01-14 (MD) hand-edited version of apsrev4-1.bst

\end{document}